# Magnonic chaotic comb

Ruitong Sun, Guanqi Ye, and Fusheng Ma*

*Key Laboratory of State Manipulation and Advanced Materials in Provincial Universities, Institute of Physics Frontiers and Interdisciplinary Sciences, School of Physics and Technology, Nanjing Normal University, Nanjing 210046, China*

*Email: phymafs@njnu.edu.cn

**ABSTRACT:**

Optical chaotic comb, possessing the key metrics of intrinsic random amplitude, phase, and frequency modulation of comb lines, emerges as a novel chaotic source in information systems for coherence tomography, parallel ranging, and secure communications. Considering the analogies between magnons and photons, the magnonic analog of optical chaotic combs is expected but not yet explored. Here, we propose a scenario of generating magnonic chaotic combs based on mode coupling mechanism in magnonic systems. Especially, we theoretically demonstrate the realization of magnonic frequency combs through three-wave mixing between ultra-strongly coupled magnons in silicon based synthetic antiferromagnet platform. It is found that the realized magnonic frequency combs can transition to chaos via various routes, *i.e.*, subcritical Hopf bifurcation, torus-doubling bifurcation, and torus breakdown. The robustness of magnonic chaotic combs is verified by characterizing the Poincaré map, the bifurcation diagrams, and the largest Lyapunov exponents. Furthermore, the unique characters of chaotic combs, perturbation hypersensitivity and noise immunity, are conceptually validated by identifying latent magnetic signal contaminated by inherent noise. Our findings provide a magnonic paradigm of chaotic dynamics in complex systems for potential applications in CMOS-integrated metrology, sensing, and communication.

# Introduction

Chaos, characterized by extreme sensitivity to initial conditions and long-term unpredictability of temporal evolution, is ubiquitous in fluids, circuits, biology, chemistry, physics, and engineering[1,2]. Chaotic dynamics has been successfully adopted for revolutionizing information technology, such as advanced sensing, secure communications, and ultra-fast random number generation[3-5]. Nowadays, frequency combs originally discovered in optics have found widespread application in frequency metrology and spectroscopy[6-8]. Very recently, a novel chaotic comb, combining the advantages of chaos and frequency combs, is eagerly pursued to obtain unique optical characters of broad bandwidth, high-level parallelization capability, high conversion efficiency, and random modulation[9-12]. For instance, the inherently chaotic nature of optical chaotic combs is harnessed for either generating massively parallel chaotic signals with orthogonal channels or implementing precise and interference-immune ranging to overcome the time and frequency congestion. Alternatively, the magnetization dynamics of magnetic systems provides an ideal platform for exploring nonlinear dynamics and chaos[13,14]. Chaotic dynamics of spin-torque nano-oscillators has been demonstrated for energy-efficient nonvolatile magnetic storage and invertible logic applications[15-21]. In the last few years, magnonic frequency combs (MFCs) have been theoretically proposed[22-28] and experimentally demonstrated[29-32] with magnon-matter interaction in the linear regime. Inspired by achievements of optical chaotic combs, analogies between magnons and photons, and inherently nonlinearity of magnetization dynamics, the magnonic analog of optical chaotic combs is expected but has yet to be realized.

Cavity magnonics is an emerging field of exploring strong magnon-matter couplings and their potential applications for quantum science and technology[33-35]. In particular, strong magnon-magnon coupling has been realized in various magnetic systems, such as bulk YIG, low-dimensional $CrCl_3$, and magnetic multilayers[36-42]. Motivated by the compatibility with CMOS technology and the miniaturization devices with nanolithography techniques, there has been an upsurge of interest in magnon-magnon coupling with magnetic multilayers sputtered on silicon substrate at room temperature[43]. Among these, synthetic antiferromagnets (SAFs) have attracted increasing attention benefiting from their agility via manipulating layer thickness and material composition[44,45]. Recently, the coupling between magnons has been successfully achieved with compensated/uncompensated SAFs in the strong, ultra-strong, and deep-strong regime, respectively[46-49]. Of particular interest is the ultra-strong coupling regime, where the coupling rate becomes comparable to the mode frequencies and overwhelms all the dissipation rates, allowing to exchange energy multiple times before losing to the environment[50-52]. Hence, the parametric instabilities of the entangled ground state in the quantum regime resulting from ultra-strong magnon-magnon coupling could have their own peculiar enchantment to explore nonlinear dynamics and chaos.

In this work, we propose a scenario of generating magnonic chaotic combs (MCCs) based on mode coupling mechanism in cavity magnonic system. The existence of MFCs and the corresponding transition to MCCs are theoretically verified with magnon-polariton, which results from ultra-strong coupling between magnons in uncompensated SAFs. The pumping induced MFCs is attributed to the nonlinear coupling between the lower and upper branches of magnon-polariton through a three-magnon process. For red-detuned, near-resonant, and blue-detuned pumping, the MFCs will evolve into MCCs via three distinct routes, *i.e.*, subcritical Hopf bifurcation, torus-doubling bifurcation, and torus breakdown, respectively. The occurrence of MCCs is further confirmed by evaluating the Poincaré map, the bifurcation diagrams, and the largest Lyapunov exponents. Moreover, the perturbation hypersensitivity and the noise immunity of MCCs are conceptually proved by extracting latent magnetic information from noisy signal.

## Results

**Scenario of MCCs**

Considering an entangled ground state resulting from ultra-strong magnon-magnon coupling as shown in Fig. 1a, the hybridized modes $f_u$ and $f_d$ of magnon-polariton would become unstable under strong pumping $h_p(t)$. Depending on the amplitude $h_p$ and frequency $f_p$ of $h_p(t)$, various nonlinear phenomena of magnetization dynamics would be approached, *i.e.*, self-oscillation, quasi-periodic oscillations, and chaos. Interestingly, MCCs, as analog of optical chaotic combs, could be realized via distinct routes. Taking the case of near-resonant pumping $f_p \approx f_u$ as an instance, the magnonic system would experience MFCs, torus-doubling bifurcation, and even MCCs with the increase of $h_p$, which are schematically shown in Figs. 1b-d. For weak $h_p$, the pumping-induced mode $f_p$ couples ultra-strongly with $f_d$ through three-magnon progress and excites both sum-frequency ($f_p + f_d$) and difference-frequency ($f_p - f_d$) modes. These secondary modes will further couple with $f_d$ to generate higher-order modes. This initiates a cascaded process of producing MFCs in the manner of $lf_p \pm nf_d$ ($l = 0, 1, 2, \ldots$; $n = 0, 1, 2, \ldots$), where $lf_p$ represents the order and center of the $l^{th}$ comb, $n$ represents the $n^{th}$ comb tooth of the $l^{th}$ comb, and $f_d$ represents the comb tooth spacing. Fig. 1b schematically shows the first three combs with $l = 0, 1, 2$, respectively. As the $h_p$ increases, the $f_d$ undergoes a torus-doubling bifurcation cascade, where the comb teeth are separated by $f_d/2, f_d/3, f_d/4$, and so on. For instance, Fig. 1c shows a period-3 torus-doubling bifurcation of $f_d$ and the resulting combs $lf_p \pm nf_d/3$. When the $h_p$ increases beyond a threshold, the torus-doubling bifurcations of $f_d$ becomes non-resolved, which results in the realization of MCCs with random modulation of comb teeth as shown in Fig. 1d. These nonlinear dynamics can be further characterized by the Poincaré map in the phase space[2], as insets shown in Figs. 1b-d. In addition, for red- and blue-detuned pumping, the magnonic system could enter into MCCs through the subcritical Hopf bifurcation and the torus breakdown routes, respectively.

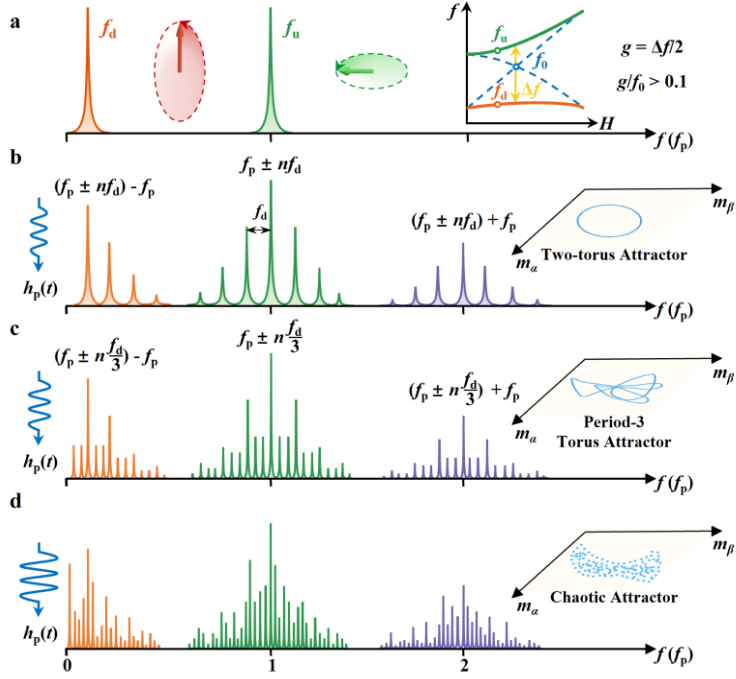

**Fig. 1 | Scenario of MCCs. a** Spectrum of hybridized modes $f_u$ and $f_d$ resulting from ultra-strongly coupled magnons. Left and middle insets present precession of magnetization corresponding to $f_u$ and $f_d$, respectively; right inset presents the frequency-field dispersion of ultra-strongly coupled magnons (solid lines) and bare magnons (dashed lines). $g = \Delta f/2$ is coupling strength and $f_0$ is coupling center frequency. Normalized coupling strength $g/f_0 > 0.1$ represents ultra-strong coupling regime. **b-d** Evolution of MFCs under near-resonant pumping $h_p(t)$ with frequency $f_p \approx f_u$ for various amplitudes $h_p$: **b** MFCs with tooth spacing $f_d$; **c** peroid-3 torus-doubling bifurcation of tooth spacing $f_d$; and **d** MCCs obtained via torus-doubling bifurcation. Insets present the corresponding Poincaré map in the phase space.

## MCCs realized with SAF

Here, we illustratively utilize the SAF system to demonstrate the proposed scenario of MCCs. The schematic diagram of the SAF is depicted in Fig. 2a, in which the antiferromagnetically coupled two ferromagnetic (FM) layers are separated by a non-magnetic layer. It is generally hard to achieve nonlinear instability of magnons in magnetic thin films sputtered on silicon, which is strongly limited by the experimentally unavailable threshold amplitude of pumping field. Notably, the presence of antiferromagnetic interlayer exchange coupling in SAF could remarkably decrease the threshold amplitude of pumping induced instability of hybridized magnons especially in the ultra-strong coupling regime[53]. Here, uncompensated SAF is adopted to prepare the hybridized ground state $f_u$ and $f_d$ from ultra-strong coupling between the in-phase mode and the out-of-phase mode. To achieve the symmetry-breaking of SAF, an intrinsic asymmetry of magnetic anisotropy is introduced between the two FM layers. The magnetization dynamics of SAF are determined by two coupled Landau-Lifshitz-Gilbert (LLG) equations[14]. Considering that the effective field $H_{\text{eff},i}$ comprises the external magnetic field $H_{\text{ext}} = H$, the anisotropy field $H_i^{\text{a}} = H_i^{\text{a}} m_i$, and the interlayer coupling field $H_{i(j)}^{\text{ex}} = H_{\text{ex}} m_{j(i)}$ ($i, j = 1, 2$ and $i \neq j$), the temporal variation of magnetization could be obtained by solving the coupled LLG equations (see details in Methods and Section S1 in Supplementary[54]):

$$\begin{cases} \dfrac{dm_i^x}{dt} = -\dfrac{\gamma}{1+\alpha_i^2}[-((m_i^y)^2 + (m_i^z)^2)(H_x + H_{\text{ex}} m_j^x)\alpha_i + (\alpha_i m_i^x m_i^y - m_i^z)(H_y + H_{\text{ex}} m_j^y) \\ \qquad + (\alpha_i m_i^x m_i^z + m_i^y)(H_z + H_i^{\text{a}} m_i^z + H_{\text{ex}} m_j^z)], \\ \dfrac{dm_i^y}{dt} = \dfrac{\gamma}{1+\alpha_i^2}[-(\alpha_i m_i^x m_i^y + m_i^z)(H_x + H_{\text{ex}} m_j^x) + ((m_i^x)^2 + (m_i^z)^2)(H_y + H_{\text{ex}} m_j^y)\alpha_i \\ \qquad + (m_i^x - \alpha_i m_i^y m_i^z)(H_z + H_i^{\text{a}} m_i^z + H_{\text{ex}} m_j^z)], \\ \dfrac{dm_i^z}{dt} = \dfrac{\gamma}{1+\alpha_i^2}[(m_i^y - \alpha_i m_i^x m_i^z)(H_x + H_{\text{ex}} m_j^x) - (\alpha_i m_i^y m_i^z + m_i^x)(H_y + H_{\text{ex}} m_j^y) \\ \qquad + ((m_i^x)^2 + (m_i^y)^2)(H_z + H_i^{\text{a}} m_i^z + H_{\text{ex}} m_j^z)\alpha_i]. \end{cases} \quad (1)$$

Where $m_{i(j)}$ represents the normalized magnetization of the two FM layers, $\gamma$ corresponds to the gyromagnetic ratio, and $\alpha_i$ denotes the Gilbert damping constant of each FM layer.

When a pumping field $h_{\text{p}}(t)$ is applied, it should be involved into $H_{\text{ext}} = H + h_{\text{p}}(t)$. Then, the temporal variation of magnetization $m_i(t)$ under pumping can be achieved from the revised Eq. (1) as:

$$\begin{cases} \dfrac{dm_i^x}{dt} = -\dfrac{\gamma}{1+\alpha_i^2}[-((m_i^y)^2 + (m_i^z)^2)(H_x + h_{\text{p}}^x(t) + H_{\text{ex}} m_j^x)\alpha_i \\ \qquad + (\alpha_i m_i^x m_i^y - m_i^z)(H_y + h_{\text{p}}^y(t) + H_{\text{ex}} m_j^y) \\ \qquad + (\alpha_i m_i^x m_i^z + m_i^y)(H_z + h_{\text{p}}^z(t) + H_i^{\text{a}} m_i^z + H_{\text{ex}} m_j^z)], \\ \dfrac{dm_i^y}{dt} = \dfrac{\gamma}{1+\alpha_i^2}[-(\alpha_i m_i^x m_i^y + m_i^z)(H_x + h_{\text{p}}^x(t) + H_{\text{ex}} m_j^x) \\ \qquad + ((m_i^x)^2 + (m_i^z)^2)(H_y + h_{\text{p}}^y(t) + H_{\text{ex}} m_j^y)\alpha_i \\ \qquad + (m_i^x - \alpha_i m_i^y m_i^z)(H_z + h_{\text{p}}^z(t) + H_i^{\text{a}} m_i^z + H_{\text{ex}} m_j^z)], \\ \dfrac{dm_i^z}{dt} = \dfrac{\gamma}{1+\alpha_i^2}[(m_i^y - \alpha_i m_i^x m_i^z)(H_x + h_{\text{p}}^x(t) + H_{\text{ex}} m_j^x) \\ \qquad - (\alpha_i m_i^y m_i^z + m_i^x)(H_y + h_{\text{p}}^y(t) + H_{\text{ex}} m_j^y) \\ \qquad + ((m_i^x)^2 + (m_i^y)^2)(H_z + h_{\text{p}}^z(t) + H_i^{\text{a}} m_i^z + H_{\text{ex}} m_j^z)\alpha_i]. \end{cases} \quad (2)$$

Without loss of generality, a bias magnetic field is applied along the *x*-axis. Based on Eq. (1), the calculated spectrum of hybridized modes $f_u$ = 7.95 GHz and $f_d$ = 0.52 GHz is shown in Fig. 2b for $H_x$ = 120 mT (see material parameters and calculated frequency-field dispersion in Section S2 in Supplementary[54]). It is observed that the chirality of $m_1$/$m_2$ is right-/left-handed for the lower hybridized mode $f_d$. In contrast, the chirality of $m_1$/$m_2$ is left-/right-handed for the upper hybridized mode $f_u$. Here, a sinusoidal pumping field $h_p(t) = h_p\sin(2\pi f_p t)$ is used in the *x-y* plane, which is tilted from *x*-axis toward *y*-axis by an angle of 45°. Hereafter, we carry out the analysis of pumping $h_p(t)$ induced nonlinear dynamics with the net magnetization $m(t) = m_1(t) + m_2(t)$, which is obtained from Eq. (2). The calculated temporal process of $m_1$ and $m_2$ are individually presented in Section S3 in Supplementary[54].

Firstly, we study the case of near-resonant pumping with $f_p$ (= 8.05 GHz) ≈ $f_u$ (= 7.95 GHz) for six typical amplitude $h_p$, which are marked by empty stars in Fig. 2c. The calculated temporal variation of $m_x$, Poincaré map, and magnonic spectra are shown in Figs. 2d-f, respectively. As $h_p$ increases, $m_x(t)$ initially exhibits periodic oscillations, then transitions to quasiperiodic behavior with anharmonic features, and eventually becomes aperiodic as displayed in Fig. 2d. The observed aperiodicity indicates the unpredictable variation of $m(t)$, which is a hallmark of chaos. These temporal features can be further understood by constructing the Poincaré map in the phase space via recording trajectory of local maxima of $m_x(t)$ as shown Fig. 2e. Initially, the trajectory exhibits as a fixed point indicating the periodic oscillations. Then, it falls on a closed curve and "doubles" into two, four, and three similar, but not wholly identical, and connected closed curves, which signify peroid-1, peroid-2, peroid-4, and peroid-3 tori, respectively. Finally, the closed curves turn to an ensemble of discrete points, *i.e.*, chaotic attractor. In addition, the magnonic spectra are also calculated as shown in Fig. 2f. For weak $h_p$, the resonantly excited mode by the pumping $f_p$ = 8.05 GHz is solely observed in the spectrum. With increasing $h_p$, MFCs are observed with comb center located at $lf_p = l \times 8.05$ GHz ($l$ = 0, 1, 2) and tooth spacing equal to $f_d$ = 0.52 GHz. Further increasing $h_p$, additional comb teeth emerge and tooth spacing switches to $f_d/2$, $f_d/4$, and $f_d/3$ sequentially. Ultimately, the comb teeth blur, expand into a disordered continuous distribution, and exhibit random modulation, which results in the realization of MCCs. Aforementioned transition from MFCs to MCCs, as we demonstrate below, can be understood as the torus-doubling bifurcation route to chaos.

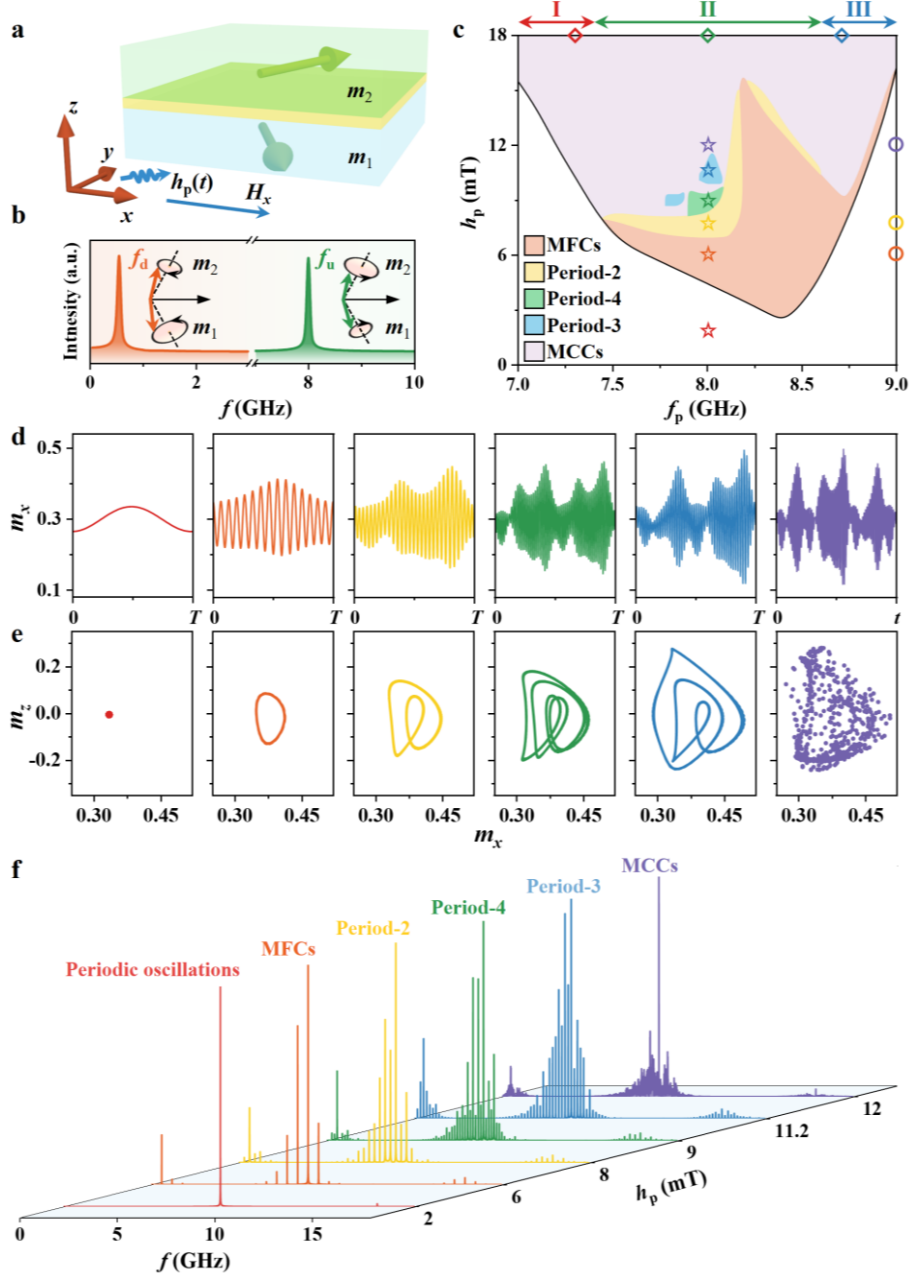

**Fig. 2 | MCCs realized with SAF. a** Schematic of SAF with magnetization configuration in the spin-canted state. **b** Calculated magnonic spectrum from $m_x$ with hybridized modes $f_u$ = 7.95 GHz and $f_d$ = 0.52 GHz for $H_x$ = 120 mT. Insets present precession of $m_1$ and $m_2$ corresponding to $f_u$ and $f_d$. **c** Phase diagram as a function of $h_p$ and $f_p$ for $H_x$ = 120 mT. Regions of MFCs, torus-doubling bifurcation, and MCCs are color coded. Three different routes to MCCs delineated by I, II, and III along $f_p$. **d** Temporal variation of $m_x$, **e** Poincaré maps of $m_x$ and $m_z$, and **f** magnonic spectra of SAF near-resonantly pumped with $f_p$ = 8.05 GHz for various $h_p$ as indicated by empty stars in **c**.

**Routes to MCCs**

Depending on $h_p$ and $f_p$, various dynamical states are observed as depicted in Fig. 2c, where the unstable regions are rigorously identified and color coded (see the magnonic spectra with various $h_p$ and $f_p$ in Section S4 in Supplementary[54]). The low boundary of color-coded region marks the threshold amplitude of $h_p$ for the onset of instability with various $f_p$. As shown in Fig. 2c, it is found that the MCCs can be realized through three qualitatively distinct routes as delineated by I, II, and III, which are corresponding to pumping frequency detunings $\Delta = f_p - f_u$. In this section, we distinguish these routes to MCCs by evaluating magnonic spectra, bifurcation diagrams, and largest Lyapunov exponents (LLEs), which are shown in Fig. 3. The bifurcation diagrams are extracted as the local maxima $m_{z\text{-max}}$ of normalized $m_z$. And the existence of chaos is justified by LLEs, which are calculated as:

$$\lambda_{\text{LLE}} = \lim_{t \to \infty} \frac{1}{t} \ln\left(\frac{|\delta m(t)|}{|\delta m(0)|}\right), \qquad (3)$$

where $|\delta m(t)|$ is the difference between the two trajectories of $m(t)$ at time $t$, and $|\delta m(0)|$ is the difference between the two close trajectories of $m(t)$ at initial time[55].

For red-detuned pumping $\Delta = -0.65$ GHz with $f_p = 7.3$ GHz, the magnonic spectra show that the resonantly excited mode becomes into disordered continuous distribution at $h_p = 10.5$ mT as displayed in Fig. 3a. This abrupt transition is also reflected in the bifurcation diagram as a sudden jumping from periodic window to chaotic window. Correspondingly, the LLEs change from negative to positive values, which indicates the appearance of chaos. This evolution suggests a subcritical Hopf bifurcation route to MCCs. For near-resonant pumping $\Delta = 0.1$ GHz with $f_p = 8.05$ GHz, it is observed from the magnonic spectra in Fig. 3b that MFCs are generated at $h_p = 4.2$ mT. Subsequently, the MFCs experience a sequence of torus-doubling bifurcations continuously from $h_p = 7.1$ mT. Finally, the MFCs blur and result in non-resolved tooth spacing at $h_p = 11.5$ mT. The bifurcations of MFCs are further revealed in the related bifurcation diagram, in which the initially stable branch successively splits into two, four, and three branches with intermittently interrupted by chaotic windows. Together with the variation of LLEs alternating between negative and positive values, the transition from MFCs to MCCs can be described as the torus-doubling bifurcation route to chaos. For blue-detuned pumping $\Delta = 0.75$ GHz with $f_p = 8.7$ GHz, three regions are observed from the magnonic spectra as shown in Fig. 3c: resonantly excited mode; MFCs; and broad distribution. The direct transition from MFCs to an elevated noise floor in spectra relates to the emergence of chaotic window in bifurcation diagram, and the occurrence of positive LLEs. This obvious transition can be explained as torus breakdown route to chaos. The temporal variation of $m_x$ and Poincaré maps of $m_x$ and $m_z$ of these routes are presented in Section S5 in Supplementary[54].

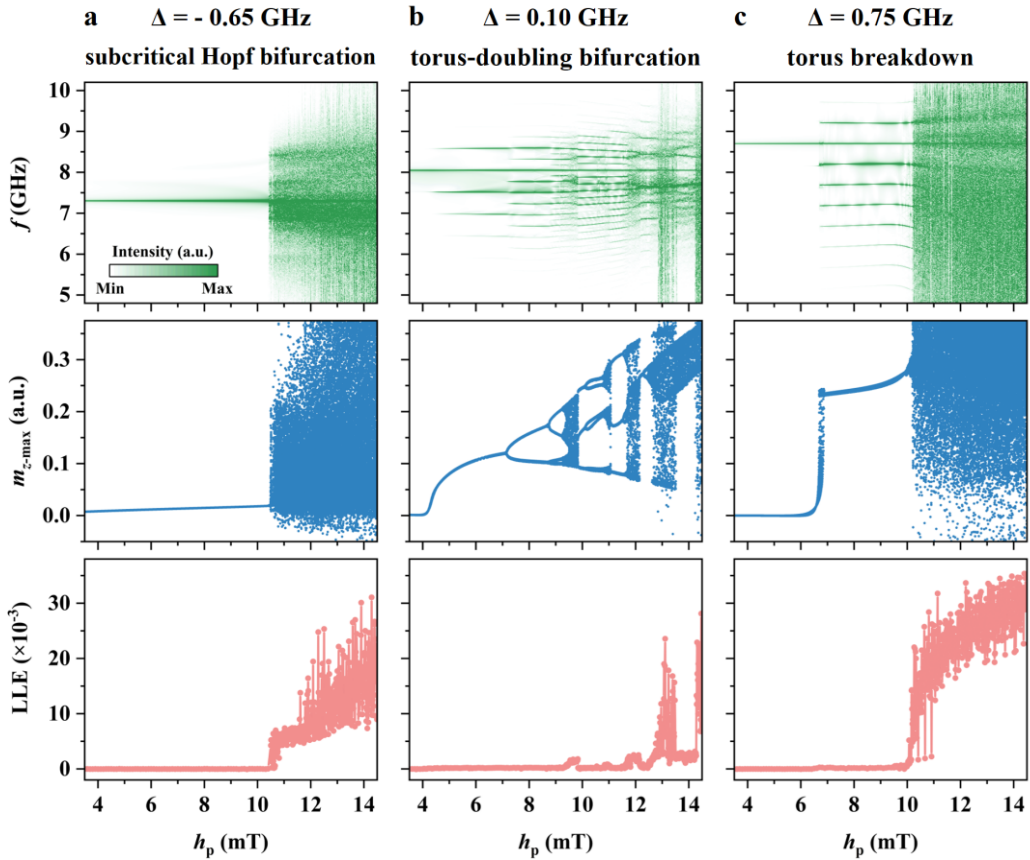

**Fig. 3 | Routes to MCCs.** Color plots of magnonic spectra, bifurcation diagrams, and largest Lyapunov exponents as a function of $h_p$ for distinct routes to MCCs: **a** subcritical Hopf bifurcation for red-detuned pumping $\Delta = -0.65$ GHz; **b** torus-doubling bifurcation for near-resonant pumping $\Delta = 0.10$ GHz; and **c** torus breakdown for blue-detuned pumping $\Delta = 0.75$ GHz. The three values of $\Delta$ are marked as empty diamonds in Fig. 2c, respectively.

On the other hand, the onset of instability is also highly dependent on frequency detuning $\Delta = f_p - f_u$ for fixed $h_p$ as shown in Fig. 2c. Fig. 4 depicts the magnonic spectra as a function of $\Delta$ for three typical $h_p = 6$, 8, and 12 mT. It is observed that the unstable regions appear in the vicinity of the crossings between the sub- and higher harmonics of fp and $f_u$ = 8.05 GHz. For instance, the 1/2, 2$^{nd}$ and 3$^{rd}$ harmonics of $f_p$ are observed for $h_p$ = 12 mT, as shown from the overall view in Fig. 4c. Here, we focus on the unstable region for $f_p$ around $f_u$, i.e., Δ ranging from -0.1 to 1.5 GHz, as the zoom-in views shown in Fig. 4. For the three $h_p$, the instability is observed in different Δ ranges: -0.4 to 0.8 GHz; -0.5 to 1.2 GHz; and -0.7 to 1.3 GHz, respectively. In these unstable regions, the spectra exhibit three distinct evolutions. For $h_p$ = 6 mT, MFCs are solely observed, and tooth spacing progressively decreases and eventually stabilizes to $f_d$ = 0.5 GHz. For $h_p$ = 8 mT, period-2 tours-doubling bifurcation of MFCs with tooth spacing as $f_d$/2 is firstly observed, and then MFCs with tooth spacing $f_d$ and MCCs appear alternately. For $h_p$ = 12 mT, the MCCs occur in most of Δ range and are interrupted by intermittent emergence of MFCs.

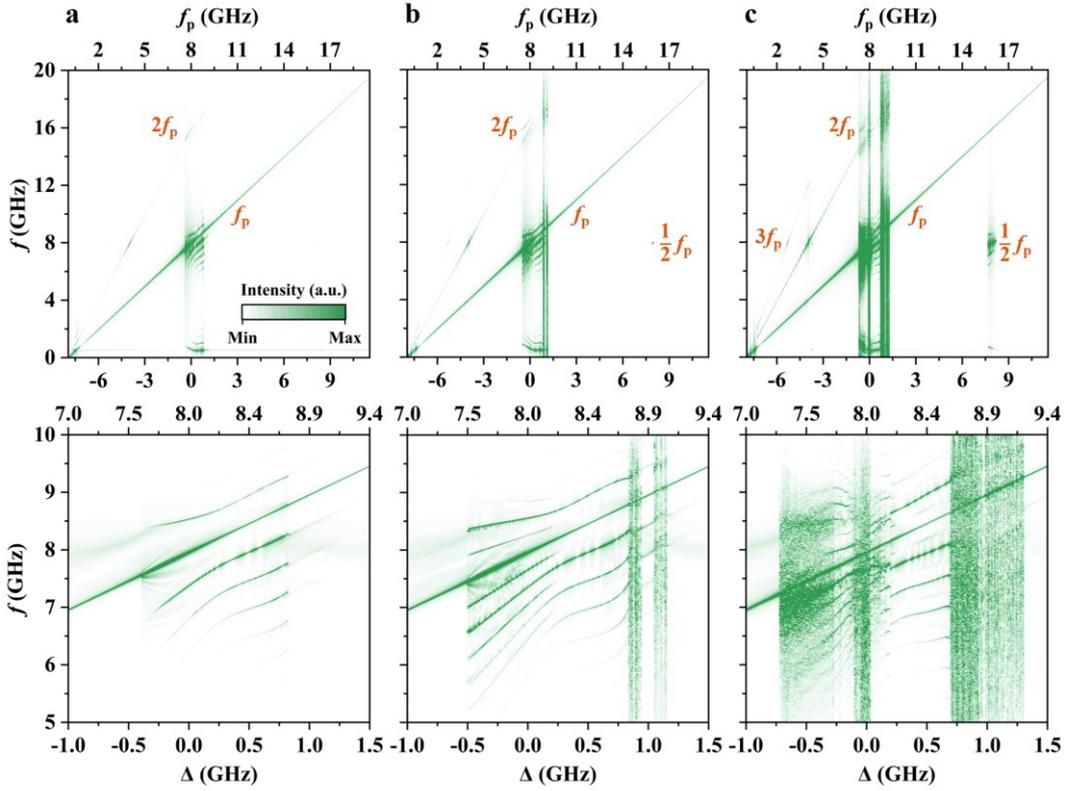

**Fig. 4 | Onset of instability by detuning Δ.** Color plots of magnonic spectra as a function of Δ for three typical pumping amplitudes: **a** $h_p$ = 6 mT; **b** $h_p$ = 8 mT; and **c** $h_p$ = 12 mT, as marked by empty circles in Fig. 2c. First row presents overall views, and second row shows the zoom-in views for $f_p$ around $f_u$.

**Noise-immune signal identification**

When a system is in the critical state, such as periodic oscillations but on the verge of transition to chaos, a small perturbation could give rise to the qualitative change. In this scenario, the pumping field $h_p(t) = h_p\sin(2\pi f_p t)$ with amplitude $h_p$ around the threshold amplitude $h_c$ for the onset of chaos drives the system into critical state serving as the reference for signal identification. If a to-be identified signal $S_0(t) = S_0\sin(2\pi f_s t + \beta)$ exists in the environment, it will modulate the $h_p(t)$ to $h_0(t) = h_p(t) + S_0(t) = h_0\sin(2\pi f_0 t + \varphi)$, where the amplitude $h_0$ and the phase $\varphi$ are as follows:

$$h_0 = \sqrt{h_p^2 + S_0^2 + 2h_p S_0 \cos[2\pi(f_s - f_p)t + \beta]}, \tag{4}$$

$$\tan\varphi = \frac{S_0 \sin[2\pi(f_s - f_p)t + \beta]}{h_p + S_0 \cos[2\pi(f_s - f_p)t + \beta]}. \tag{5}$$

When $h_p \gg S_0$, the variation of phase could be negligible, *i.e.*, $\varphi \sim 0$. As indicated by Eq. (4), the $h_0$ will periodically vary around $h_c$ within the range $h_p \pm S_0$ with $\Delta f = |f_s - f_p|$. Consequently, the regular and chaotic oscillations occur alternately, which results in intermittent chaos with a periodicity of:

$$T = \frac{1}{\Delta f}. \tag{6}$$

Therefore, intermittence of our proposed MCCs can be utilized to identify latent magnetic signals. For instance, Fig. 5a shows the magnonic spectra for various $h_p$ with $f_p = 7.5$ GHz. An edge of MCCs is observed at $h_p = 8.6$ mT. Fig. 5b shows the temporal variation of $m_x$ corresponding to the MFCs and MCCs, respectively. If there is a signal $S_0(t)$ in the environment, it can be identified by tuning the $f_p$. Fig. 5c schematically shows the case for $S_0 = 0.5$ mT and $f_s = 7.5$ GHz. The intermittent character can be found from the $m_x(t)$ and magnonic spectra with the temporal interval $T = 100$ ns by tuning $f_p$ from 7.5 to 7.51 GHz as shown in Fig. 5c. As derived from Eq. (6), the $\Delta f$ is calculated as 0.01 GHz, corresponding to $f_s = 7.5$ or 7.52 GHz. By considering the case of $f_p = 7.49$ GHz, the $f_s$ can be confirmed as 7.5 GHz as shown in Section S6 in Supplementary[54]. Furthermore, the magnetic signal can still be identified by analyzing the existence of intermittent MCCs even the $S_0(t)$ is buried in environmental noise. As an instance, Fig. 5d shows the case of $S_0(t)$ buried in a Gaussian noise $\sigma\xi(t)$, which satisfies $\langle\xi(t)\xi(t')\rangle = \delta(t - t')$ with a standard deviation $\sigma = 0.5$ mT.

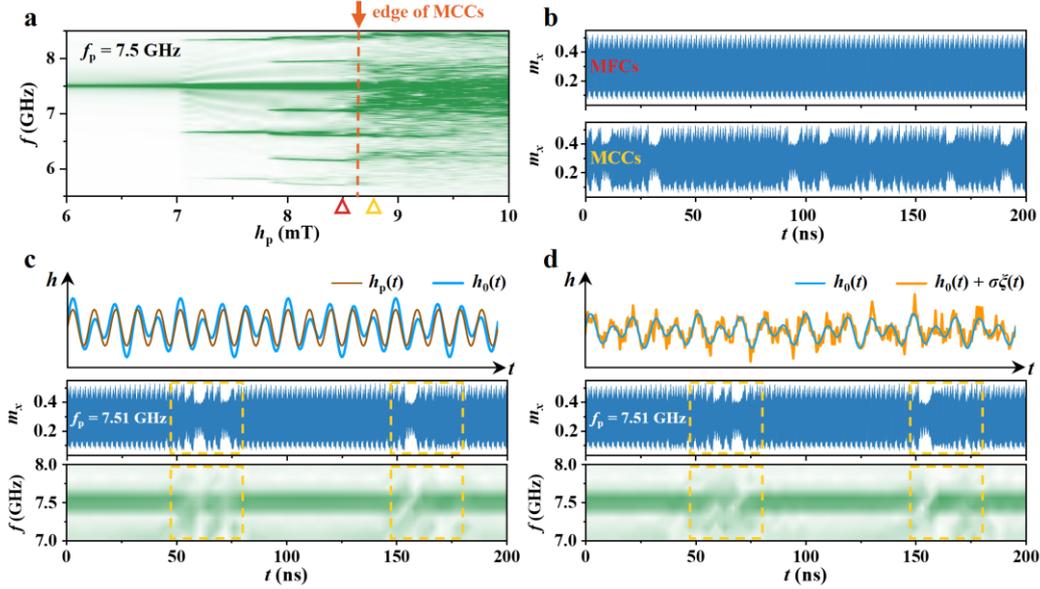

**Fig. 5 | Proof-of-concept of noise-immune signal identification. a** Color plots of magnonic spectra for $f_p$ = 7.5 GHz. **b** $m_x(t)$ corresponds to two distinct states, *i.e.*, MFCs and MCCs, marked by red and yellow empty triangles in **a**, respectively. Schematic of $h(t)$, temporal variation of $m_x$, and magnonic spectra obtained from Short-Time Fourier Transform for: **c** $h_0(t) = h_p(t) + S_0(t)$; and **d** $h_0(t) + \sigma\xi(t)$. Dashed boxes indicate the occurrence of intermittent MCCs.

## Discussion

In summary, we propose a scenario of generating chaotic combs based on ultra-strongly coupled magnons and demonstrate it using synthetic antiferromagnet. Depending on the pumping frequency detuning, the transition to MCCs is observed via three distinct routes, *i.e.*, subcritical Hopf bifurcation, torus-doubling bifurcation, and torus breakdown. We further verify the MCCs by characterizing Poincaré map, bifurcation diagrams, and largest Lyapunov exponents. Moreover, we conceptually validate a noise-immune signal identification by utilizing the chaotic nature of MCCs. It is also found that the existence of MCCs is significantly dependent on the bias magnetic field and magnon-magnon coupling strength, see Sections S7 and S8 in Supplementary[54], respectively.

Recently, ultra-strong magnon-magnon coupling have been demonstrated experimentally in compensated ferrimagnet[38], antiferromagnets[41,42], and SAFs[49]. In addition, the threshold amplitude of pumping for our proposed scenario of MCCs is within the current experimental reach. Therefore, experimental demonstration of our findings is fully feasible. Our findings pave the way for further exploring nonlinear dynamics with magnonic system such as turbulence, stochastic resonance, synchronization of chaos, and quantum chaos, as well as the potential applications for high-precision metrology, sensing, and communication.

**Methods**

The magnetization dynamics of SAF are governed by coupled LLG equations:

$$\frac{d\boldsymbol{m}_i}{dt} = -\gamma \boldsymbol{m}_i \times \boldsymbol{H}_{\text{eff},i} + \alpha_i \boldsymbol{m}_i \times \frac{d\boldsymbol{m}_i}{dt}, \quad (7)$$

where $\boldsymbol{m}_i$ ($i = 1, 2$) represents the normalized magnetization of the two FM layers, $\gamma$ corresponds to the gyromagnetic ratio, and $\alpha_i$ denotes the Gilbert damping constant of each FM layer. $\boldsymbol{H}_{\text{eff},i}$ is the effective field, including the external magnetic field $\boldsymbol{H}_{\text{ext}}$, the anisotropy field $\boldsymbol{H}_i^{\text{a}} = H_i^{\text{a}} \boldsymbol{m}_i$, and the interlayer coupling field $\boldsymbol{H}_{i(j)}^{\text{ex}} = H_{\text{ex}} \boldsymbol{m}_{j(i)}$ ($i, j = 1, 2$ and $i \neq j$). To separate the time-dependent terms, the coupled LLG equations are reformulated in matrix form as follows:

$$\begin{bmatrix} \frac{dm_i^x}{dt} \\ \frac{dm_i^y}{dt} \\ \frac{dm_i^z}{dt} \end{bmatrix} = -\gamma \begin{bmatrix} m_i^x \\ m_i^y \\ m_i^z \end{bmatrix} \times \begin{bmatrix} H_{\text{eff},i}^x \\ H_{\text{eff},i}^y \\ H_{\text{eff},i}^z \end{bmatrix} + \alpha_i \begin{bmatrix} m_i^x \\ m_i^y \\ m_i^z \end{bmatrix} \times \begin{bmatrix} \frac{dm_i^x}{dt} \\ \frac{dm_i^y}{dt} \\ \frac{dm_i^z}{dt} \end{bmatrix}. \quad (8)$$

Collecting all time-dependent terms to the left-hand side, Eq. (8) is expressed in the equivalent form:

$$\begin{bmatrix} \frac{dm_i^x}{dt} \\ \frac{dm_i^y}{dt} \\ \frac{dm_i^z}{dt} \end{bmatrix} = -\gamma \begin{bmatrix} 1 & \alpha_i m_i^z & -\alpha_i m_i^y \\ -\alpha_i m_i^z & 1 & \alpha_i m_i^x \\ \alpha_i m_i^y & -\alpha_i m_i^x & 1 \end{bmatrix}^{-1} \begin{bmatrix} m_i^y H_{\text{eff},i}^z - m_i^z H_{\text{eff},i}^y \\ m_i^z H_{\text{eff},i}^x - m_i^x H_{\text{eff},i}^z \\ m_i^x H_{\text{eff},i}^y - m_i^y H_{\text{eff},i}^x \end{bmatrix}. \quad (9)$$

Expanding the matrix equation into a set of differential equations, the coupled LLG equations are recast as follows:

$$\begin{cases} \dfrac{dm_i^x}{dt} = -\dfrac{\gamma}{1+\alpha_i^2}[-((m_i^y)^2 + (m_i^z)^2)\alpha_i H_{\text{eff},i}^x + (\alpha_i m_i^x m_i^y - m_i^z)H_{\text{eff},i}^y + (\alpha_i m_i^x m_i^z + m_i^y)H_{\text{eff},i}^z], \\ \dfrac{dm_i^y}{dt} = \dfrac{\gamma}{1+\alpha_i^2}[-(\alpha_i m_i^x m_i^y + m_i^z)H_{\text{eff},i}^x + ((m_i^x)^2+(m_i^z)^2)\alpha_i H_{\text{eff},i}^y + (m_i^x - \alpha_i m_i^y m_i^z)H_{\text{eff},i}^z], \\ \dfrac{dm_i^z}{dt} = \dfrac{\gamma}{1+\alpha_i^2}[(m_i^y - \alpha_i m_i^x m_i^z)H_{\text{eff},i}^x - (\alpha_i m_i^y m_i^z + m_i^x)H_{\text{eff},i}^y + ((m_i^x)^2 + (m_i^y)^2)\alpha_i H_{\text{eff},i}^z]. \end{cases} \quad (10)$$

The temporal variation of the $\boldsymbol{m}_i$ can be obtained by solving Eq. (10), which is further utilized for characterizing the two fundamental magnon modes, *i.e.*, in-phase mode and out-of-phase mode. For symmetrical SAF, the in-phase mode and out-of-phase mode will intersect at a crossing point. Once the symmetry of SAF is broken, the coupling between the in-phase mode and out-of-phase mode will occur, forming an anti-crossing gap. By introducing an intrinsic asymmetry of magnetic anisotropy between two FM layers, the ultra-strong magnon-magnon coupling can be realized. Under a bias magnetic field and a pumping field, the $\boldsymbol{H}_{\text{ext}}(t)$ is adopted as $H_x \boldsymbol{e}_x + h_p(\boldsymbol{e}_x + \boldsymbol{e}_y)\sin(2\pi f_p t)$, and the Eq. (10) is revised as:

$$\begin{cases} \dfrac{dm_i^x}{dt} = -\dfrac{\gamma}{1+\alpha_i^2}[-((m_i^y)^2 + (m_i^z)^2)(H_x + h_p(t) + H_{\text{ex}} m_j^x)\alpha_i + (\alpha_i m_i^x m_i^y - m_i^z)(h_p(t) + H_{\text{ex}} m_j^y) \\ \qquad\qquad + (\alpha_i m_i^x m_i^z + m_i^y)(H_i^a m_i^z + H_{\text{ex}} m_j^z)], \\ \dfrac{dm_i^y}{dt} = \dfrac{\gamma}{1+\alpha_i^2}[-(\alpha_i m_i^x m_i^y + m_i^z)(H_x + h_p(t) + H_{\text{ex}} m_j^x) + ((m_i^x)^2+(m_i^z)^2)(h_p(t) + H_{\text{ex}} m_j^y)\alpha_i \\ \qquad\qquad + (m_i^x - \alpha_i m_i^y m_i^z)(H_i^a m_i^z + H_{\text{ex}} m_j^z)], \\ \dfrac{dm_i^z}{dt} = \dfrac{\gamma}{1+\alpha_i^2}[(m_i^y - \alpha_i m_i^x m_i^z)(H_x + h_p(t) + H_{\text{ex}} m_j^x) - (\alpha_i m_i^y m_i^z + m_i^x)(h_p(t) + H_{\text{ex}} m_j^y) \\ \qquad\qquad + ((m_i^x)^2+(m_i^y)^2)(H_i^a m_i^z + H_{\text{ex}} m_j^z)\alpha_i]. \end{cases} \quad (11)$$

By measuring time in units of $(\gamma M_s)^{-1}$, we obtain the dimensionless Eq. (11) and integrate it through a standard fourth-order Runge-Kutta integration scheme with a fixed time step $\Delta t = 2$ ps, where $M_s$ is the saturation magnetization. The used magnetic parameters of SAF are: $M_s = 6 \times 10^5$ A/m; $\alpha_1 = \alpha_2 = 0.01$; $H_1^a = 139$ mT; $H_2^a = -500$ mT; $H_{\text{ex}} = -200$ mT; and $|\gamma|/2\pi = 0.027$ GHz/mT. The calculated coupling strength is $g/f_0 = 0.5$ entering into the ultra-strong coupling regime. The detailed theoretical calculation is presented in Section S1 in Supplementary[54].

## Data availability

All relevant data are available from the corresponding authors on request.

## Acknowledgements

The authors thank Peiqing Tong for helpful discussions on chaos theory; Yu Zhang for suggestions on Poincaré map; and Junwen Sun for advice on Runge-Kutta integration scheme. This work was supported by the National Key Research and Development Program of China (Grant No.2023YFF0718400) and the National Natural Science Foundation of China (Grant Nos. 12474119 and 12074189)

## Author contributions

F. Ma conceived the research and supervised the study. R. Sun performed the theoretical calculations. R. Sun, G. Ye, and F. Ma performed the analysis of results and the writing of the manuscript. All authors have read and approved the final manuscript.

## Competing interests

The authors declare no competing interests.